\documentclass[12pt]{iopart}

\usepackage{graphicx}
\usepackage{siunitx}
\usepackage{todonotes}
\usepackage[super]{nth}
\usepackage{amsfonts,amssymb}

\usepackage[backend=biber,sorting=none]{biblatex}
\begin{document}

\title{Reminding Forgetful Organic Neuromorphic Device Networks}

\author{Daniel Felder}
\address{DWI - Leibniz Institute for Interactive Materials,  Forckenbeckstrasse 50, Aachen, DE\\
AVT.CVT - Chair of Chemical Process Engineering, RWTH Aachen University, Forckenbeckstrasse 51, Aachen, DE}

\author{Katerina Muche}
\address{AVT.CVT - Chair of Chemical Process Engineering, RWTH Aachen University, Forckenbeckstrasse 51, Aachen, DE}

\author{John Linkhorst}
\address{AVT.CVT - Chair of Chemical Process Engineering, RWTH Aachen University, Forckenbeckstrasse 51, Aachen, DE}

\author{Matthias Wessling}
\address{DWI - Leibniz Institute for Interactive Materials, Forckenbeckstrasse 50, Aachen, DE\\
AVT.CVT - Chair of Chemical Process Engineering, RWTH Aachen University, Forckenbeckstrasse 51, Aachen, DE}
\ead{manuscripts.cvt@avt.rwth-aachen.de}

\begin{abstract}
Organic neuromorphic device networks can accelerate neural network algorithms and directly integrate with microfluidic systems or living tissues. 
Proposed devices based on the bio-compatible conductive polymer PEDOT:PSS have shown high switching speeds and low energy demand.
However, as electrochemical systems, they are prone to self-discharge through parasitic electrochemical reactions. 
Therefore, the network's synapses forget their trained conductance states over time. 
This work integrates single-device high-resolution charge transport models to simulate entire neuromorphic device networks and analyze the impact of self-discharge on network performance.
Simulation of a single-layer nine-pixel image classification network commonly used in experimental demonstrations reveals no significant impact of self-discharge on training efficiency. 
And, even though the network's weights drift significantly during self-discharge, its predictions remain 100\% accurate for over ten hours. 
On the other hand, a multi-layer network for the approximation of the circle function is shown to degrade significantly over twenty minutes with a final mean-squared-error loss of 0.4.
We propose to counter the effect by periodically reminding the network based on a map between a synapse's current state, the time since the last reminder, and the weight drift. 
We show that this method with a map obtained through validated simulations can reduce the effective loss to below 0.1 even with worst-case assumptions. 
Finally, while the training of this network is affected by self-discharge, a good classification is still obtained. 
Electrochemical organic neuromorphic devices have not been integrated into larger device networks. 
This work predicts their behavior under nonideal conditions, mitigates the worst-case effects of parasitic self-discharge, and opens the path toward implementing fast and efficient neural networks on organic neuromorphic hardware. 
\end{abstract}

\vspace{2pc}

\section{Introduction}
Neuromorphic devices can accelerate large neural networks to unlock new applications in natural language processing \cite{thoppilan2022lamda}, image generation\cite{saharia2022photorealistic} and many more. 
At the same time, the energy efficiency of neuromorphic devices allows the integration of small neural networks into low-powered devices for edge applications \cite{christensen20222022}. 
Organic neuromorphic devices allow, furthermore, a direct integration into electrolytic systems and can directly communicate with interactive materials such as artificial cells through ion channels, microgels through ion release and sensing \cite{han2020microfabricated, chen2020materials}, and living tissues such as nerve cells \cite{keene2020biohybrid}.
These properties open a pathway towards intelligent systems that combine sensing, processing, and memory \cite{biomorphic,kaspar2021rise}.

Recently several organic neuromorphic device concepts have been presented. 
Van de Burgt et al. demonstrated electrochemical random access memories (ECRAMs) \cite{van2017non} that have been adapted to work at elevated temperatures \cite{melianas2020temperature} or to be screen printable \cite{liu2019fully}.
Lee et al. propose a similar device with a nanofiber-based channel and ultra-fast response times \cite{lee2021nanofiber}. 

ECRAMs based on PEDOT:PSS and a solid-state electrolyte are among the very promising candidates. 
They feature high write linearity for efficient neural network training algorithms, low energy demand, and decoupled read and write operations through a separate gate electrode. 
However, they have been shown to exhibit self-discharge, which makes a neural network's weights gradually drift toward zero over time.
While physical mitigation strategies have been proposed, perfect self-discharge-free devices are difficult to produce, especially for scaled-down versions and devices that are supposed to interact with other interactive material systems.
This weight drift with long timescales can be useful in some scenarios like eligibility traces in reinforcement learning \cite{demiraug2021pcm}.
For conventional, pre-trained neural networks, it results, however, in a gradual degradation in prediction performance that needs to be precisely understood.

This work uses comprehensive numeric models to understand the effect of self-discharge on neural network performance and enable its mitigation  through efficient algorithm-hardware co-design. 
It is shown that some networks are highly vulnerable to self-discharge, while others are completely immune.
Furthermore, algorithmic countermeasures are proposed to stabilize vulnerable networks for extended amounts of time. 

\begin{figure}
    \centering
    \includegraphics[width=\textwidth]{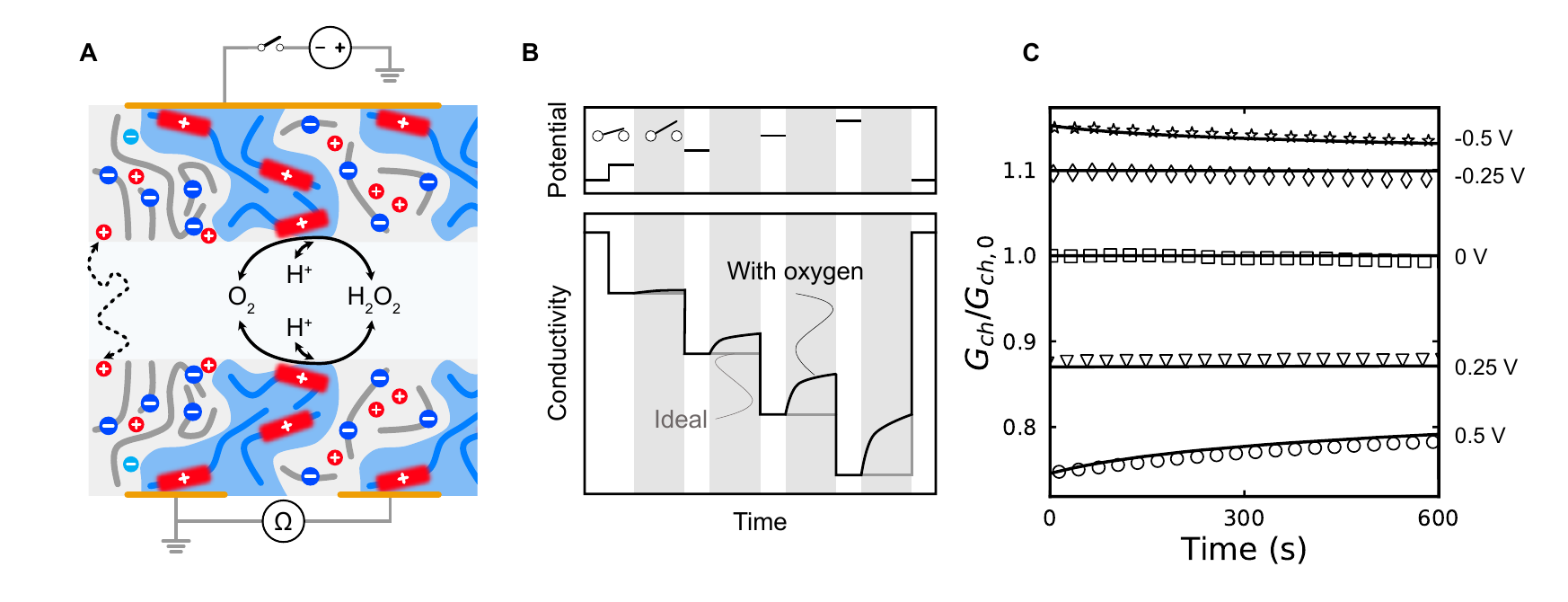}
    \caption{Organic neuromorphic device structure, functionality, and limitations. \textbf{(A)} Three-terminal neuromorphic device consisting of a PEDOT:PSS gate (top), electrolyte (center), and PEDOT:PSS channel (bottom). \textbf{(B)} Device programming through voltage pulses on the gate. The resulting conductivity change is impermanent in the presence of depolarizing impurities such as oxygen and falls back to the initial state. \textbf{(C)} Conductance change over time for devices programmed to an initial gate voltage on the right. Symbols represent experimental observations and lines are predicted by the device model. A-C are partially redrawn from Felder et al. \cite{felder2022coupled}}
    \label{fig:model}
\end{figure}

\section{Methods}
This work aggregates physically accurate single-device models into multiple device arrays. 
Each array represents a neural network layer and is simulated transiently for backpropagation and inference.

\subsection{Single-device model}
The single-device model is based on Nernst-Planck-Poisson two-phase charge transport modeling of an electronically conductive PEDOT phase and an ionically conductive PSS and electrolyte phase. 
Both are coupled capacitively with an approach first presented by Tybrandt et al. \cite{tybrandt2017chemical}.
Additionally, self-discharge is described through an electrochemical shuttle reaction modeled with a Butler-Volmer kinetic, and water dissociation is described by mass action law kinetics. 
Figure \ref{fig:model}A visualizes the single device with the modeled effects. 
Writing and reading are decoupled during neuromorphic operations. 
Therefore, lateral ion transport in the channel can be neglected, and the modeling domain reduced to a 1D line between the gate and drain electrode.
These assumptions also hold when devices are integrated into crossbars and are valid as long as low read potentials and sufficient relaxation times between reads and writes are used.
The single-device model's equations, characteristics, and validation were described in detail in a previous publication \cite{felder2022coupled}.

Figure \ref{fig:model}B visualizes single-device programming.
Increasing potentials at the gate electrode result in lower conductances in the channel, while decreasing potentials increase the conductance.
The devices can be switched reproducibly between many discrete states \cite{van2017non}.
However, in the presence of impurities such as ambient oxygen, these states quickly decay towards the initial device conductance. 
The effect is caused by an electrochemical shuttle reaction between the two PEDOT:PSS layers.
One of the identified reactions is a hydrogen peroxide shuttle \cite{bamgbopa2021modelling}.
As the potential increases, the reaction and, therefore, the state decay gets exponentially faster. 
This results in highly stable states close to the initial conductance and much less stable states for larger or smaller conductances.

Figure \ref{fig:model}C shows a comparison of real-world experimental data of self-discharge with the applied model. 
States with an initial gate potential around \SI{\pm 0.25}{\volt} decay insignificantly over \SI{10}{\min}, while states with an initial potential around \SI{\pm 0.5}{\volt} decay by up to \SI{12}{\percent} of the total range in the same time. 
The data highlight an additional nonideality that has to be considered for some devices. 
Nonlinear conductance tuning causes the states in the figure to lie closer together for negative potentials. 
This is caused by crowding in the density of states that results in a lowered electron mobility and is accounted for in the model through an experimentally obtained correlation.
Mitigation strategies through partial imine dededoping have been successfully employed in literature \cite{van2017non, keene2019mechanisms, keene2020enhancement}.
Therefore, ideal linear devices are assumed unless otherwise noted in the following.

\subsection{Crossbar model}
\begin{figure}
    \centering
    \includegraphics[width=\textwidth]{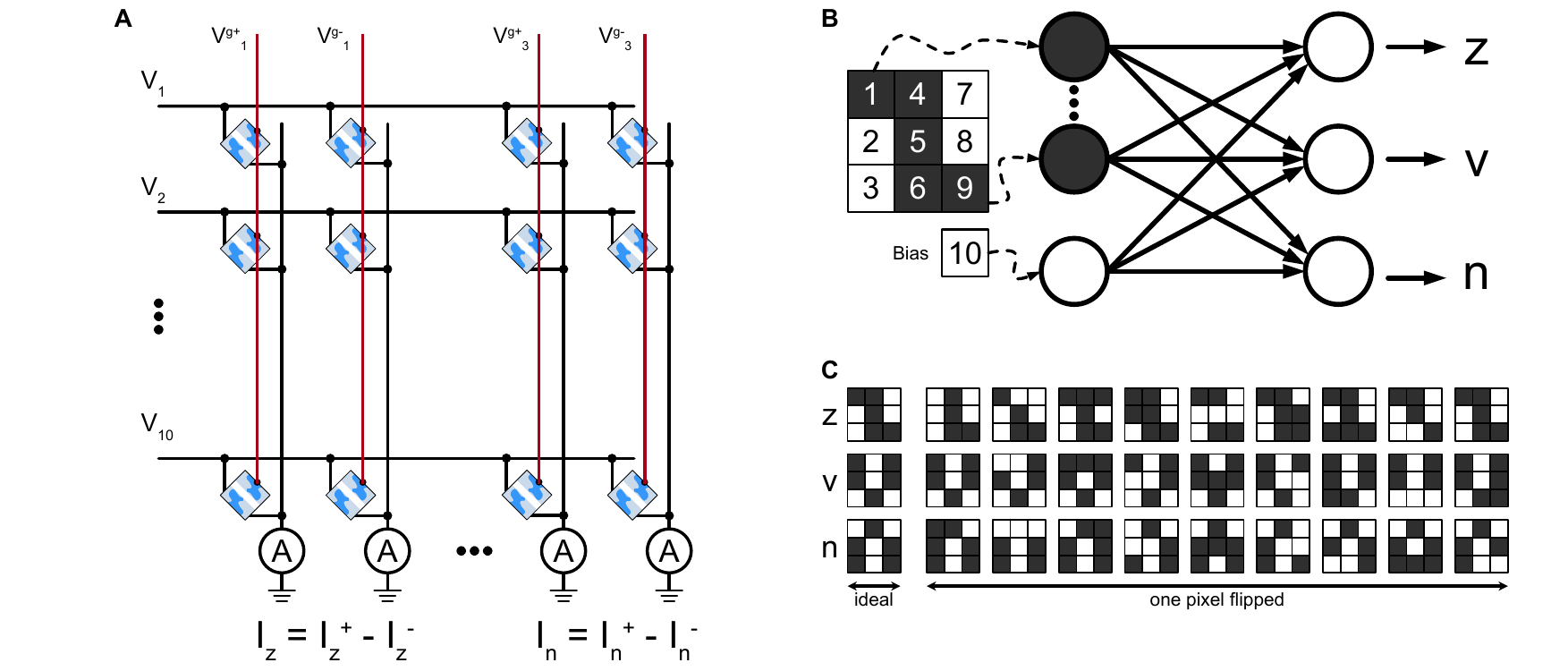}
    \caption{Neural network hardware implementation. \textbf{(A)} Crossbar of organic neuromorphic devices with differential synapses. Input potentials are applied on the left, gate potentials on the top, and resulting currents are read on the bottom. \textbf{(B)} Example network for the crossbar in A that classifies 3x3 pixel images into the categories z, v, and n. \textbf{(C)} Input data to the example network. Ideal pictures on the left, noisy pictures on the right.}
    \label{fig:concept}
\end{figure}

To manufacture large scale device networks, single devices are commonly integrated into crossbar structures. 
For two-terminal devices, a set of parallel leads connects to the devices' bottom terminal with another set of leads perpendicular to the first connecting the top electrode.
This results in a configuration where every device can be addressed individually for programming and up to a full row can be programmed at the same time.
For electrochemical neuromorphic devices with three terminals, such an architecture can be adapted to include a third set of leads.
Figure \ref{fig:concept}B visualizes this concept with a crossbar built with differential artificial synapses. 
Differential synapses compensate for fluctuations in the devices' characteristics during fabrication by combining two devices and subtracting the devices' resulting currents from each other.
During inference, each device in the differential synapse receives the same voltage inputs. 
During training, the devices are programmed into opposite directions with the same but inverted gate voltages.
To map device conductance to neural network weights, they are normalized with a reference initial conductance $G^{0,\text{ref}}$.
Each device's gate voltage range is limited to between \SI{-0.5}{\volt} and \SI{0.5}{\volt}.
Therefore, weights are capped at -1 and +1.
The neural network algorithms can partially compensate for this limitation through a weight-scale factor $\beta \geq 1$. 
Networks with bounded weights have previously been studied to limit overfitting at the cost of reduced learning capability \cite{liao2004neural}. 
Therefore, differential weights $W_{ij}$ with the following formula are created.
\begin{equation}
    W_{ij} = \beta \cdot \frac{G_{ij}^+ - G_{ij}^-}{G^{0,\text{ref}}}
\end{equation}
where $G_{ij}^\pm$ is the devices' conductance. 

When single-device models are placed in a crossbar configuration, their potential interactions must be considered.
The devices themselves do not directly influence each other. 
However, the length of the leads to the power source and transimpedance amplifier changes from device to device resulting in variable parasitic resistances. 
It has been shown that the channel conductance of the devices can be finely tuned through partial imine dedoping of PEDOT \cite{van2019mechanism}. 
We, therefore, assume that it is possible to build devices with significantly higher channel resistance than the leads and expect the lead resistances to be negligible during read operations. 
For writing, the gate to channel impedance cannot be easily modified, but programming pulses can be chosen long enough to get close to steady-state independently of parasitic resistances.
Therefore, the crossbar is modeled as a network of single-device models with ideal connections.

\subsection{Inference}
During a forward pass through each network layer, inputs are encoded as continuous voltages $V_j$ and scaled to the range of \SI{\pm 0.1}{\volt} by multiplication with $\gamma = \SI{0.1}{\volt}$
Limiting of input voltages of consecutive layers is achieved through the corresponding activation functions. 
With Kirchhof's law the following output current results for each neuron in each layer
\begin{equation}
    I_i^\pm = \sum_{j=1}^N V_j G_{ij}^\pm 
\end{equation}
where $N$ is the number of neurons in the previous layer.
The currents are converted back into voltages by a transimpedance amplifier with a rate of $\beta \cdot (G^{0,\text{ref}})^{-1}$, subtracted and passed through a hardware implementation of the $f(x) = \gamma\cdot \text{tanh}(\gamma^{-1} \cdot x)$ activation function.
An ideal implementation of the activation function is assumed to focus on the unique properties of organic neuromorphic devices as synapses.
tanh is chosen to maintain comparability with Prezioso's work \cite{prezioso2015training}, and because it is symmetric and zero-centered.
The minimal current range of the transimpedance amplifier is given by $I_\text{max} = N \cdot \SI{0.1}{\volt} \cdot G|_{V_g = \SI{-0.5}{\volt}}$.
The voltages at the output layer are finally scaled back up by a factor of $\gamma^{-1}$ to obtain real-valued outputs.

\subsection{On-device training}
On-device training allows the neural networks to adapt to manufacturing variability between the different neuromorphic devices and, therefore, partially compensate it.
Furthermore, on-device training benefits from the same highly parallel multiply and accumulate operations as the inference step.
Here, a modified, discrete Delta Rule approach known as Manhattan Rule is applied following Prezioso et al.'s demonstration of training on metal oxide devices \cite{prezioso2015training,zamanidoost2015manhattan}.

Weights are initialized by random, normally distributed \SI{1}{\second} pulses between \SI{-0.5}{\volt} and \SI{0.5}{\volt}.
\SI{1}{\second} pulses were previously shown to be long enough for capacitive charging processes to reach a near-equilibrium state and short enough to suppress electrochemical reactions.
If weight-scaling is applied through $\beta > 1$, the pulses are divided by $\beta$ before being constrained to the maximum gate potential of \SI{\pm 0.5}{\volt}.
Weight updates are performed through \SI{100}{\milli\second} pulses with positive or negative potentials versus the device's current open circuit potential (OCP).
In Manhattan Rule training, each update pulse has the same fixed height $\eta$ equivalent to a learning rate.
Mean-squared-error (MSE) is used as loss function.
For the output layer, weight increments $\Delta_{ij}(n)$ for each data point $n$ are calculated as
\begin{eqnarray}
    \delta_i(n) = \left[t_i(n) - y_i(n)\right] \cdot \text{tanh}'(a_i(n))\\
    \Delta_{ij}(n) = \delta_i(n)x_j(n)
\end{eqnarray}
where $x$ are the input data, $t$ the target values, $a$ the layer's outputs before the activation function, and $y$ after the activation function.
$\Delta_{ij}(n)$ are computed for one $n$ at a time and aggregated. 
After all inputs have been applied, the synaptic weights are updated by the Manhattan Rule. 
\begin{equation}
    \Delta W_{ij} = \eta \cdot \text{sgn} \sum_{n=1}^{N} \Delta_{ij}(n)
\end{equation}

For multi-layer networks, the backpropagation algorithm is applied for the preceeding layers. 
\begin{eqnarray}
    \delta_{i,l}(n) = (\delta_{l+1}(n) \cdot \beta W_l^\intercal)_i \cdot \text{tanh}'(a_{i,l}(n))\\
    \Delta_{ij,l}(n) = a_{i,l-1}(n) \delta_{j,l(n)}
\end{eqnarray}
where $l$ denotes the layer number from input towards output layer. 
Here, the backpropagation step $\delta_{l+1} \cdot \beta W_l^\intercal$ is applied through reversal of the crossbar.
Crossbar reversal is described in detail in the Supporting Information section S1.

\subsection{Self-discharge simulation}
After virtual on-device training, a model for every neuromorphic device in every layer is obtained with an individual conductance state and concentration distribution.
During the self-discharge simulation, these device models are advanced through time in open circuit mode. 
No electrical current flows between the source or drain and the gate electrode.
Changes result purely from the electrochemical shuttle reaction and the resulting ion movement. 
Channel conductances are calculated from the channel's electronic charge carrier concentration and monitored over time. 
Every \SI{20}{\second} the networks' loss functions are evaluated to monitor prediction performance decline.

\subsection{Device reminders}
In contrast to, for example, PCM devices \cite{ambrogio2019reducing}, the conductance drift of organic devices is a complex function of the device's state and, therefore, more difficult to compensate.
However, with the presented model, each device's conductance decline can be precisely predicted depending on its history and initial state. 
These predictions are used to design reminder pulses to reverse the effects of self-discharge.
For an efficient implementation, a calibration map between the initial synapse weight, the time of self-discharge and the current weight is created. 
Data is produced by simulating synapses for a range of starting weights $w_\text{start} \in [0, 0.1, 0.2, ..., 0.9, 1]$ for \SI{20000}{\second}.
Since differential synapses are symmetric, no negative starting weights are required.
The data is then transformed into a map between the current weight, the time of self-discharge and the resulting delta in the weight's value.
This map is interpolated in 2D to obtain weight deltas for arbitrary points within the data limits. 

\subsection{Read-error effect estimation}
To estimate the effect of read-errors on inference and backpropagation, weight sets with normally distributed read errors are created during each read operation.
It is assumed that the transimpedance amplifiers do not have a systematic error and the mean of the distribution is zero, the standard deviation is based on Keene's observations \cite{keene2019mechanisms}. The normal distribution is chosen as default lacking more experimental data.
\begin{eqnarray}
    W_{ij,\text{read}} = W_{ij,\text{ideal}} + \text{err}_{ij}\\
    \text{err}_{ij} \sim \mathcal{N}(0, 0.0025^2)
\end{eqnarray}
During backpropagation, the weights are sampled with new errors for each inference and backpropagation step.

\begin{figure}
    \centering
    \includegraphics[width=\textwidth]{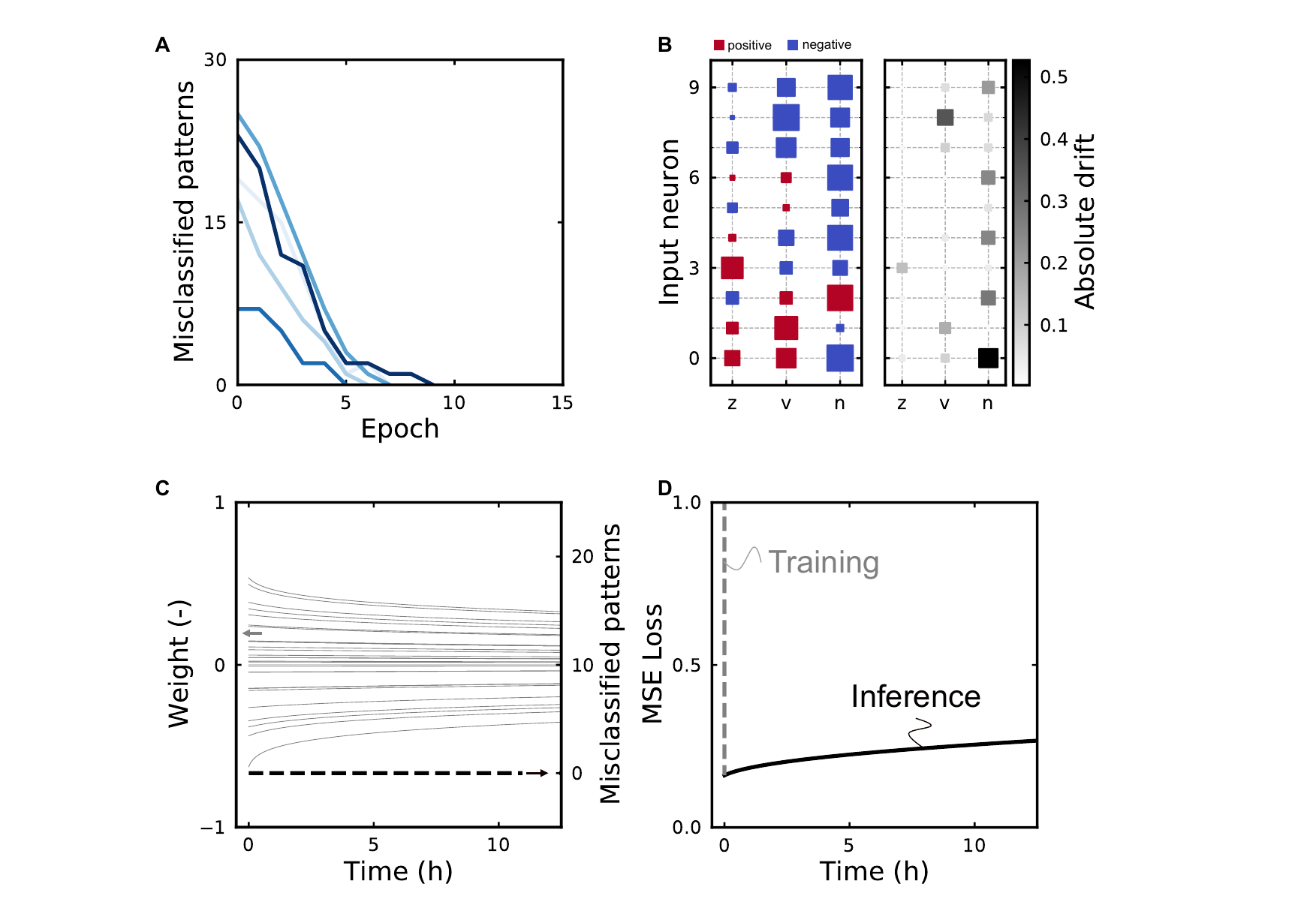}
    \caption{Behavior of a single layer network on organic neuromorphic hardware. \textbf{(A)} Training progress of the network with device drift, bounded weights, and Manhattan Rule updates. \textbf{(B)} Exemplary network weights after training and total drift of each weight over \SI{10}{\hour}. \textbf{(C)} Weight evolution over time (left) and prediction performance of the network (right) during self-discharge. \textbf{(D)} Evolution of the loss function over time during self-discharge.}
    \label{fig:slperceptron}
\end{figure}

\section{Results}
The effect of forgetful neuromorphic devices is analyzed for two example neural networks.
One is a single-layer perceptron designed to recognize 3x3 pixel images as letters z, n, and v. 
Similar networks are often used for the demonstration of real-world neuromorphic devices \cite{giannopoulos20188, prezioso2015training}.
The other is a multi-layer network to approximate the circle function based on points in a 2D space. 
For both, good prediction performance is obtained with simulated on-device training. 
While the single-layer perceptron seems completely immune to the effects of self-discharge, the multi-layer network is affected strongly.

\subsection{Single-layer image classification}
The single-layer image classification network schematically drawn in Figure \ref{fig:concept}A is designed to recognize 3x3 pixel representations of the letters z, v, and n. 
The nine pixels are encoded to $-1$ for black and $1$ for white. 
As an additional \nth{10} input $-1$ is passed into the device network to replace the output neurons' biases by the weights between the \nth{10} input and the corresponding output neuron without further adjustment to the crossbar structure. 
The network is trained on the images drawn in Figure \ref{fig:concept}C. 
Additionally to the perfect representations on the left, noise is introduced by flipping one pixel in each additional image \cite{prezioso2015training}.
In view of the small size of the dataset, no distinction between training and test set is made.

Figure \ref{fig:slperceptron}A shows multiple simulated on-device training cycles with self-discharge simulated at ambient oxygen concentrations. 
Each cycle starts from a different set of normally-distributed random weights but is otherwise identical. 
Consistently, less than ten epochs are required to reach perfect classification of the training set. 
This is comparable to training cycles on metal oxide devices presented by Prezioso et al. where between 5 and 35 epochs are required \cite{prezioso2015training}.
The resulting weights (Figure \ref{fig:slperceptron}B) are, for a randomly picked run, relatively equally distributed with synapses connecting to n having slightly larger values on average. 
Drift for the first \SI{10}{\hour} is strongest for the large weights around $\pm 0.6$ and reaches up to \SI{30}{\percent} of the total range. 
For smaller weights up to $\pm 0.1$, drift is inconsequential with \SI{1}{\percent} of the total range.
As Figure \ref{fig:slperceptron}C, however, shows, even after \SI{10}{\hour} and significant weight drift, no classification errors are introduced.
At the same time, the network's loss rises from 0.14 to 0.24.

We hypothesize that this highly stable network behavior is caused by two factors. 
First, the network only has a single layer and nonlinearity is only introduced at the output. 
Every weight drifts in the same direction towards zero, but the relative order of weight sizes is preserved.
Therefore, no single output neuron is favored, and the predictions for each image are unlikely to change. 
Second, the network has many active synapses connected to each output neuron which results in large input values to the activation function of each output neuron. 
Even after a long self-discharge duration, the low gradient in the tanh activation function for large values allows only small increases in the mean-squared-error loss.

\begin{figure}
    \centering
    \includegraphics[width=\textwidth]{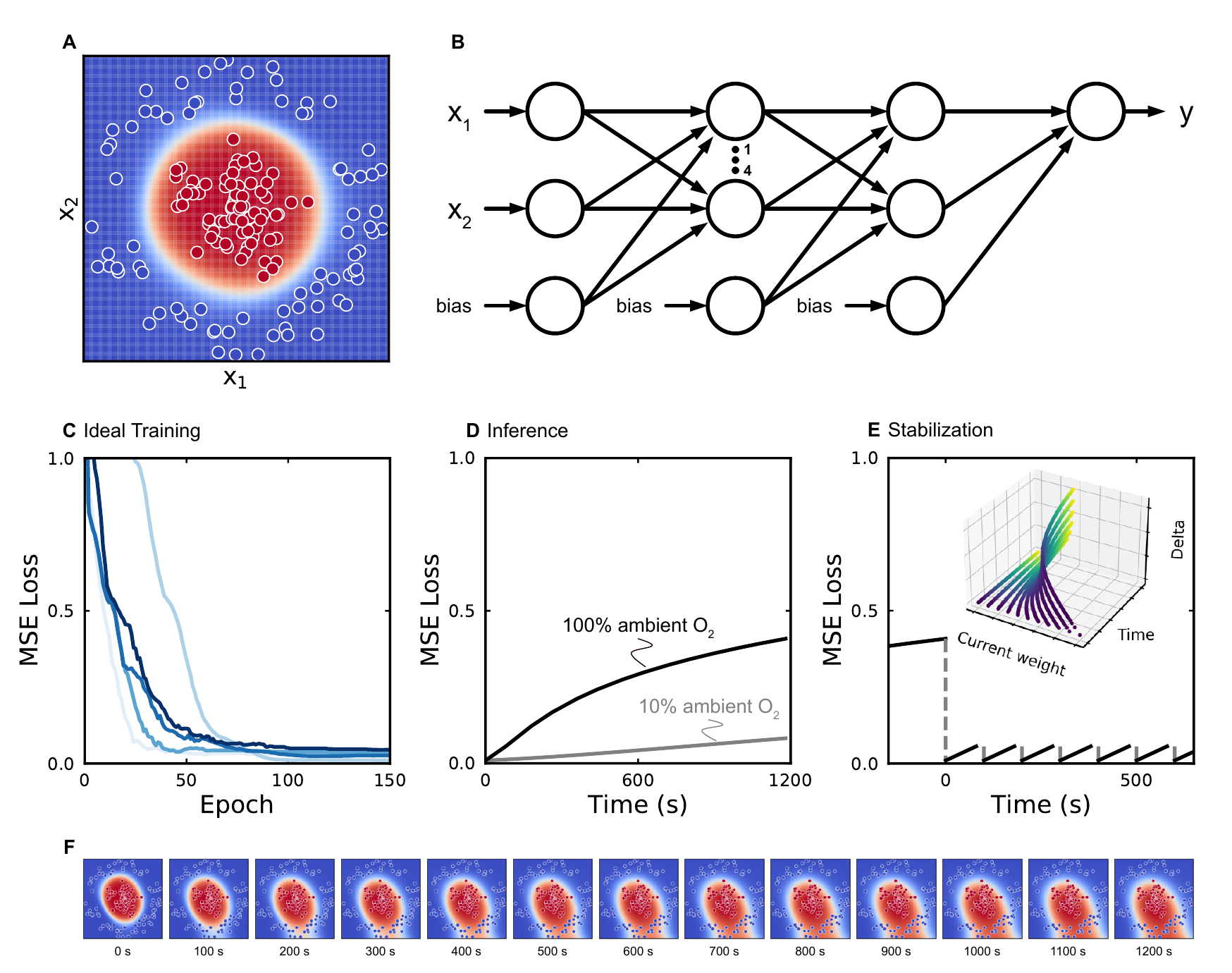}
    \caption{Behavior of a multi-layer network on organic neuromorphic hardware. \textbf{(A)} Input data (points) inside the circle (red) and outside (blue). The background shows network predictions after training of areas inside the circle (red) and outside (blue). \textbf{(B)} Structure of the network with input coordinates $x_1$ and $x_2$ and prediction $y$. \textbf{(C)} Network training with bounded weights and Manhattan Ruel but without self-discharge. \textbf{(D)} Evolution of the trained network's loss over time during self-discharge for \SI{10}{\percent} and \SI{100}{\percent} of ambient oxygen. \textbf{E)} Network loss with enabled reminder pulses. The inset shows the map between the current device state and the required reminder. \textbf{(F)} Visual evolution of the networks decision boundary over time during self-discharge.}
    \label{fig:circlemlp}
\end{figure}

\subsection{Multi-layer regression network}
As second example, a multi-layer network for the approximation of the circle function $r^2=x_1^2+x_2^2$ is analyzed (similar to \cite{tfplayground}). 
The network includes nonlinear effects through tanh activation function and two hidden layers.
It has an even shape with two input neurons, four and two neurons in the hidden layers and one output neuron as drawn in Figure \ref{fig:circlemlp}B. 
Thus, a stronger reaction to self-discharge is expected compared to the single-layer perceptron.

Data points are randomly generated in 2D ($x_1$ and $x_2$) and classified by whether they lie inside or outside a circle with the radius \num{0.5}, see Figure \ref{fig:circlemlp}A.
No data points with a radius between 0.5 and 0.7 are generated to allow a clearer separation of the two classes. 
Initially, training without self-discharge with the Manhattan Rule is attempted and results in good classification rates after 60 to 150 epochs as shown in Figure \ref{fig:circlemlp}C.
Here, a step function learning rate scheduler is applied to increase convergence speed. 
\begin{equation}
    \eta = 0.03 \cdot 0.7 \left\lfloor  \frac{1 + \text{epoch}}{20}\right\rfloor \cdot \beta^{-1}
\end{equation}

During inference, a drastic effect of self-discharge is shown over a time frame of \SI{20}{\min}.
For ambient oxygen concentrations, an increase in the network's loss function from 0.01 to 0.41 is observed and drawn in Figure \ref{fig:circlemlp}D.
For a \SI{90}{\percent} reduced oxygen concentrations, the loss still rises to 0.08.
Figure \ref{fig:circlemlp}F visualizes the change in the network's predictions by drawing the decision boundary as a red and blue background behind the training data points in red (inside the circle) and blue (outside the circle) in the foreground. 
After \SI{10}{\min} a large portion of the outside points is misclassified as inside the circle. 

\subsection{Reminding forgetful device networks}
As the previous section shows, drift in forgetful neuromorphic devices can quickly degrade the prediction performance of neural networks.
With the single device charge transport model, this drift can be precisely predicted and reversed through engineered reminder pulses (see methods section).
Figure \ref{fig:circlemlp}E shows in the inset the reset map between the current synapse weight, the time of self-discharge, and the delta pulse that is required to reset each synapse to its original state.
A quantitive version of the inset can be found in the supporting information Figure S2.
Since differential synapses are used, each synapse's behavior is perfectly symmetric around zero, even if a single device's behavior is not. 
Therefore, only the positive weight range is calculated and shown in the figure. 
For weights close to zero or after very short self-discharge times, almost no corrections are necessary.
For times of \SI{20000}{\second} and weights approaching 1, the delta approaches 0.63.
The main graph shows how the engineered reminder pulses affect the network's MSE loss function over time. 
A device network trained to approximate the circle function that self-discharged for \SI{20}{\min} is taken as base state. 
At \SI{0}{\second} the first reminder is applied to the network and results in a decrease in loss from 0.41 to 0.01.
Applying the same procedure every \SI{100}{\second} continuously keeps the network's loss below 0.06 even with worst-case assumptions about device stability.
While only virtual device reminders were tested in this work, the used models have been shown to validate well against real-world devices. 
Therefore, the presented approach is expected to transfer to real-world devices. 
The additional energy consumption by the reminders can be expected to be well below \SI{10}{\percent} of the ideal network's consumption. A detailed estimation is provided in the supporting information, section S5.

\begin{figure}
    \centering
    \includegraphics[width=\textwidth]{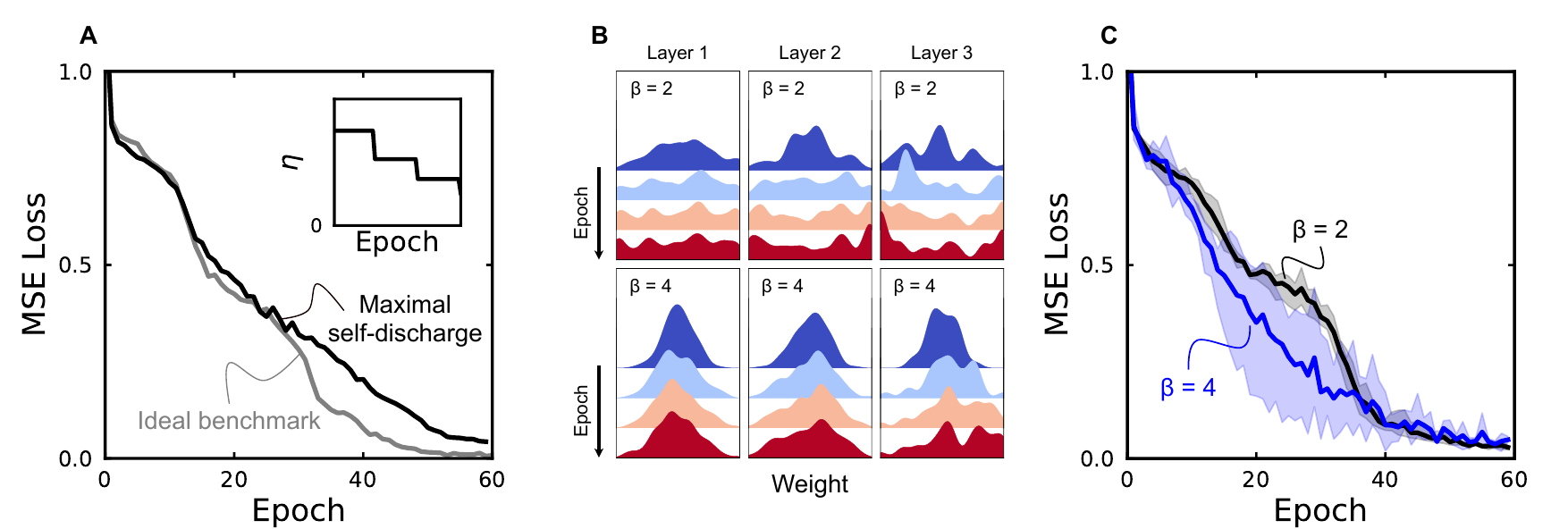}
    \caption{Effect of self-discharge during training. \textbf{(A)} Comparison of the training progress for devices without self-discharge (gray) and devices with maximal self-discharge (black). \textbf{(B)} Distribution of weights during training for epochs 0, 20, 40, and 60 for different $\beta$. The distributions are sampled from 10 randomly initialized training runs. \textbf{(C)} Estimation of the influence of device read-error for different $\beta$. Three runs per $\beta$ are drawn with line representing the mean loss, and the area the min and max loss.}
    \label{fig:o2train}
\end{figure}

\subsection*{Training forgetful device networks}
Given the vulnerability of the multi-layer network towards self-discharge during inference, the influence of self-discharge and nonlinearity during training is evaluated separately.
Figure \ref{fig:o2train}A shows the loss function during training with and without self-discharge. 
Here, the same initialization is chosen for the weights to allow a one-to-one comparison. 
During the initial epochs up until epoch 25, only small deviations in the training progress can be observed. 
After epoch 25, the effect of self-discharge becomes visible, with the ideal approaching 0.002 until epoch 60 in the reference case. 
With self-discharge, a limit of 0.014 is approached after 60 epochs for the same initial weights.
Figure \ref{fig:o2train}B shows the network's weight distribution for epoch 20, 40, and 60 during nonideal training.
Especially in the output layer (layer 3), weight values are pushed towards large values.
These large values cause a large vulnerability towards self-discharge also during training. 
Adapting the weight scaling for larger possible values by doubling $\beta$ to 4 allows the device conductance values to remain closer to the initial state and, therefore, remain more stable. 
Here, a trade-off has to be made. 
Using a smaller part of the conductance range reduces the effects of self-discharge. 
At the same time, the effects of read and write errors start to become more dominant. 
In online training, write errors are eventually canceled out.
Read errors, however, can a) affect the final accuracy of the network and b) affect backpropagation performance. 
Figure \ref{fig:o2train}C, shows simulations with a commonly assumed \cite{keene2019mechanisms} read error of \SI{0.5}{\percent}  for $\beta=2$ and $\beta=4$.
The increased stochasticity affects increase initial convergence but also limits the final accuracy for $\beta=4$.
Here, the accuracy gains from reduced self-discharge are completely neutralized.
Therefore, training can, in this case, not be enhanced through $\beta$-scaling. 
Instead, more synapses and a modified network structure would be required to decrease the loss further.
However, the resulting final loss still leads to good classification in this case.

Training with nonlinear devices, on the other hand, results in large residual errors during training, see Supporting Information Figure S3.
Even with differential devices, update symmetry is not perfect, and the total conductance range available is reduced.
While asymmetry can be compensated algorithmically \cite{gokmen2020algorithm}, state-of-the-art devices achieve highly linear conductance switching through partial imine dedoping and should be preferred \cite{van2019mechanism}. 

\section{Discussion}
Many neuromorphic devices are vulnerable to self-discharge and weight drift. 
At the same time, conventional neural network algorithms are often not well-suited to compensate these effects. 
We have demonstrated a model-based approach to quickly assess these effects for organic, electrochemical device networks and develop mitigation strategies. 

The approach was applied to two example networks. 
The single-layer perceptron for 3x3 image classification showed good stability both during training as well as inference for extended self-discharge times of up to \SI{10}{\hour}. 
We hypothesize that a combination of a large number of active synapses per output neuron and no effective nonlinearity in the network leads to the network's resilience against weight drift. 

Training a three-layer network on the circle function reveals a much stronger influence of weight drift. 
Training of the network requires less than \SI{3}{\minute} for 60 epochs with \SI{100}{\milli\second} update pulses.
Therefore, only small weight changes through self-discharge are expected. 
At the same time, on-device training is able to react to and correct the weight's drift.
Still, a reduced convergence speed and a larger final loss of 0.014 instead of 0.002 without self-discharge are observed.
There exists potential to optimize the weight distribution towards smaller conductance ranges and, therefore, reduce self-discharge that affects large weights most strongly. 
However, a trade-off with the hardware's read-accuracy has to be considered. 
During inference, weight drift has a significant effect and degrades the network's prediction performance to a loss of 0.41 after \SI{20}{\minute} for devices with uninhibited self-discharge. 
We propose a method of stabilizing these networks by periodic reminder pulses. 
These pulses can be effectively designed with the device model and require only the current device state and the time since the last reminder pulse as input to correct device drift. 
In the example network, pulses once per \SI{100}{\second} are shown to consistently keep the loss below 0.06 without impacting the network's speed or energy efficiency. 
This is an important result, as it shows that the limitations of organic neuromorphic hardware can be efficiently mitigated through algorithm-hardware co-design.

Algorithm-hardware co-design can unlock further improvements through network optimization. 
E.g., as the number of weights in a network increases, their magnitude can be decreased \cite{giannopoulos20188,gupta2015deep,courbariaux2014training}; in the extreme, even binarized networks are possible \cite{courbariaux2016binarized}. 
Additionally, effective always-on, unsupervised learning methods can eliminate the need for artificial device reminders \cite{khacef2019self,diehl2015unsupervised} and even turn self-discharge into a useful platform feature to take full advantage of the devices' properties.

We conclude that neural networks on forgetful neuromorphic hardware can be effectively implemented but require adjusted algorithms whose development is aided by model-based design.

\section{Data availability statement}
The data that support the findings of this study are available upon reasonable request from the authors.

\section{Acknowledgements}
Matthias Wessling acknowledges DFG funding through the Gottfried Wilhelm Leibniz Award 2019 (WE 4678/12-1).

\printbibliography

\end{document}